\magnification=\magstep1
\baselineskip=12pt plus 1pt minus 1pt
\parskip=3pt 

\centerline
{\bf CONSTRAINING DARK ENERGY WITH THE DARK ENERGY SURVEY:}
\centerline{\bf THEORETICAL CHALLENGES}

%\citestyle{nature}
%\bibliographystyle{apj}

\smallskip
\noindent
James Annis,
Sarah Bridle,
Francisco J. Castander,
August E. Evrard, 
Pablo Fosalba,
Joshua A. Frieman,
Enrique Gazta{\~n}aga,
Bhuvnesh Jain,
Andrey V. Kravtsov, 
Ofer Lahav,
Huan Lin,
Joseph Mohr,
Albert Stebbins,
Terence P. Walker,
Risa H. Wechsler, 
David H. Weinberg,
Jochen Weller 
\medskip
\noindent
{\bf 
1. Introduction
}

The Dark Energy Survey (DES) will use a new imaging camera on the
Blanco 4-m telescope at CTIO to image 5000 square degrees of sky in
the South Galactic Cap in four optical bands, and to carry out repeat
imaging over a smaller area to identify and measure lightcurves of
Type Ia supernovae.$^1$ The main imaging area overlaps the planned
Sunyaev-Zel'dovich survey of the South Pole Telescope.  The idea
behind DES is to use four distinct and largely independent methods to
probe the properties of dark energy: baryon oscillations of the power
spectrum, abundance and spatial distribution of clusters, weak
gravitational lensing, and Type Ia supernovae.  This white paper
outlines, in broad terms, some of the theoretical issues associated
with the first three of these probes (the issues for supernovae
are mostly different in character), and with the general task of
characterizing dark energy and distinguishing it from alternative
explanations for cosmic acceleration.  
A companion white paper
discusses the kind of numerical simulations and other theoretical
tools that will be needed to address the these issues and to create mock
catalogs that allow end-to-end tests of analysis procedures.  Although
we have been thinking about these problems in the specific context of
DES, many of them are also relevant to other planned dark energy
studies.  

\medskip
\noindent
{\bf 
2. Baryon Oscillations
}

   Oscillations of the coupled photon-baryon fluid in the early
Universe imprint a ``standard ruler'' on the matter power spectrum.
The length of this standard ruler can be calibrated using cosmic
microwave background (CMB) anisotropy measurements, in particular from
the Planck experiment.  Measurements of the galaxy power spectrum in
the transverse and line-of-sight directions then yield values of the
angular diameter distance $d_A(z)$ and Hubble parameter $H(z)$,
respectively.$^2$ This experiment can use the broad-band shape
of the power spectrum in addition to the oscillation wavelengths
themselves, at the price of somewhat stronger model dependence.

At a fundamental level, the critical issue for this approach to dark
energy studies is the effect of non-linearity, redshift-space
distortions, and complex galaxy bias on the galaxy power spectrum.
Simulations that incorporate all of these effects show that the basic
oscillation signal survives on large scales, and that the wavelengths
of the oscillation peaks are
close to those predicted by linear theory.$^3$ However, it is not
yet clear that departures from linear theory are much smaller than the
desired statistical errors.  Work is needed to address this question
at higher precision, and to understand the sensitivity of the
observable baryon oscillation scale to the assumed galaxy formation
physics, e.g., to the parameters of semi-analytic galaxy formation
models used to place galaxies in dark matter simulations.  These
studies must also investigate the dependence of the observable
oscillation scale on the cosmological model itself, since the
non-linear and galaxy bias effects could vary with the dark energy
parameters that one is trying to extract.

   Including the broad-band shape of the power spectrum allows tighter
dark energy constraints by making more complete use of the data,
but the broad-band shape of the spectrum may be more susceptible
than the baryon oscillation scales to weakly scale-dependent galaxy
bias.  Even if scale-dependent bias is present, it is possible 
that halo occupation modeling of small and intermediate scale clustering 
can pin down this scale-dependence, allowing recovery of the 
underlying linear theory power spectrum shape without reliance
on a detailed model of galaxy formation.  This approach requires development 
and testing to determine its efficacy and robustness.

At a more practical and survey-specific level, the key ``theoretical''
issue is how to best extract the desired signal from the data.  In the
case of DES, galaxy redshifts will be estimated photometrically, with
expected errors of $\Delta z \sim 0.05-0.1$.  A simple approach is to bin
galaxies by photometric redshift and study just the angular power
spectrum, but a more complete likelihood approach that uses the
distribution of photometric redshift errors should, in principle,
perform better by making more complete use of the available
information.  It is also important to understand the effects of
drifts in photometric calibration zero-points and of 
systematic errors in photometric redshifts.
In particular, it will be necessary not only to quantify
the photometric bias and scatter, but also to control the
{\it uncertainties} in the bias and scatter using careful
calibration with large spectroscopic training sets.$^{1}$
There has already been a substantial effort
to quantifying the potential impact of these effects in the case of
DES,$^{4}$ but further work is needed on methods to
identify such effects if they are present and to mitigate their
effects if they are.  The signal-extraction issue is also connected to
the physics issues discussed above, since it may be possible to
parameterize and marginalize over the most uncertain aspects of the
predictions.$^3$

\medskip
\noindent
{\bf 3. Clusters}

The cluster mass function (defined here as the number of clusters
per unit mass per unit solid angle per unit redshift) depends on dark 
energy through the growth factor and the volume element.  
The primary method of constraining
these with a cluster sample is to have some observable proxy for
cluster mass and then to measure the mass (proxy) function over a
range of redshifts to constrain dark energy.  In DES, the main proxies
will be the SZ decrement and optical richness; for other surveys,
X-ray observables will also be crucial.  With the deep imaging from
DES, it is also possible to measure, via weak lensing, accurate
cluster-mass correlation functions (i.e., population-averaged mass
profiles) for ensembles of systems selected based on one of these
proxies.$^5$ 

    The key theoretical issue is making predictions for the actual
observables rather than for idealized mass functions.  Doing so
directly requires simulations that incorporate gravity, gas dynamics,
radiative cooling, star formation, and feedback, all in a realistic
cosmological context.  The necessary physics is sufficiently
complicated that we probably won't have convincing direct predictions
of the mass-observable relations at the required level of accuracy
(i.e., percent or better) in the near future.  Rather, the best prospects
for exploiting clusters arise from calibrating the mean mass-observable
relations using observations.$^6$ 
This can be done either using observables from
the primary sample, such as the clustering of the clusters themselves,
the clustering of galaxies in the sample, or the cluster-mass
correlations estimated from weak lensing,
or it can be done using the properties of a moderate
number of objects that have been studied very well with different
techniques (adding, for example, galaxy spectroscopic data, X-ray data,
or deeper SZ or weak lensing maps).

    Simulations play two key roles in this process.  First, they
can be used to understand the magnitude and form of scatter about the
mean relations and the expectations for redshift evolution and
cosmology dependence.  The inferences about dark energy have an
important dependence on the scatter about mean relations,$^{6,7}$
but the dependence is less direct than that on the mean relation
itself, and simulations can more reliably predict the variation from
system to system and the effect of small parameter changes than the
absolute values of observables.  Calculations using different
simulation techniques and different physical assumptions will be
needed to understand the remaining uncertainties.  The second
essential role of simulations is to calibrate biases in observational
analyses and to understand how the intrinsic mass-observable relation
will propagate to observed quantities.  Detailed simulations can be
used to create mock sky surveys of clusters, including projection
effects, beam smearing, and so forth, and to make accurate predictions
in observable space that can be directly compared with data.

    The relation between the weak lensing signal and the projected mass
distribution is well understood.  Accurate measurements of the
tangential shear profiles of ensembles of clusters may therefore prove
to be a robust way of constraining dark energy, since the only
important observational biases are the effects of scatter in moving
clusters into or out of the selected ensemble and of miscentering the
clusters. This approach requires further investigation to understand
its sensitivity to dark energy parameters and its likely systematic
uncertainties; the latter is being explored in detail with lower
redshift data from the Sloan Digital Sky Survey.

\medskip
\noindent
{\bf 
4. Weak Lensing
}

Cosmic shear depends on dark energy through the distance-redshift relation,
space curvature,
and the evolution of the linear growth factor.  
Because cosmic shear directly probes the mass distribution, 
which is dominated by dark matter, it is less sensitive to baryonic physics
than the galaxy clustering or cluster abundance probes discussed above.
However, while the baryon component is sub-dominant, 
baryonic effects on the mass power spectrum can be as large
as the differences between interesting
dark energy models on the small scales where the 
signal-to-noise ratio of the measurements will be high.
A key physical issue for dark energy studies with cosmic shear
is high precision calculation of the non-linear 
matter power spectrum, including the baryonic effects. 
Even for pure collisionless dark matter, numerical simulation 
predictions and analytical models have not reached the
level of accuracy required for the next generation of surveys.$^8$
Fully assessing the baryonic effects requires simulations that 
include a dissipative baryonic component and star formation and
feedback physics that yields a realistic mass fraction in
galaxies.

A rapidly growing area of current theoretical research is the
development of methods that use weak
lensing measurements, including higher-order statistics and
correlations with the galaxy distribution
and the CMBR anisotropy, to separate the effects of geometry, curvature, and
gravitational growth.$^9$
These new theoretical ideas will increase the
robustness of weak lensing measurements of dark energy parameters, once
they are reliably put into practice through simulation and real world
observations.

The presence of intrinsic alignments in the galaxy population
is a potential source of systematic bias in cosmic shear measurements,$^{10}$
and more work is needed on theoretical predictions and empirical
estimates of these alignments and on data analysis methods
that can mitigate their impact.

At a more practical level, much of the present ``theoretical'' effort
is focused on finding the best ways
of analyzing and calibrating the imaging data, to both reduce and
estimate systematic uncertainties.  For example, recent improvements
in the method of tracking the anisotropy of the point-spread function
have led to dramatic reductions in systematic errors (identifiable
via ``B-mode'' shear polarization) from cosmic shear measurements
using the BTC and Mosaic II imagers on the CTIO Blanco telescope.
The estimation and correction of systematic errors 
is an area of much active research using both empirical and 
formal approaches.$^{11}$ Analysis methods have already progressed to
the point that it is possible to create mock galaxy catalogs for
lensing analysis that include gravitational shear as well as models of
PSF patterns. With such catalogs, lensing measurement methods can be
tested and the level of systematic error estimated in advance of the
survey to refine observing strategy. 

\medskip
\noindent
{\bf 5. Characterizing Dark Energy and Alternatives}

In the simplest cosmological models with dark energy, the universe is
spatially flat, and the dark energy parameters to be constrained are
$\Omega_{\rm DE}=1-\Omega_m$ and the equation of state parameter
$w = p/\rho$.  In the observational white paper on DES, we present
error forecasts on $w$ for each of the four probes:
$\Delta w = 0.04$, 0.11, 0.02, and 0.02 from Type Ia supernovae,
angular galaxy clustering, cluster abundances, and cosmic shear,
respectively, assuming Planck priors (see reference [1], 
Table 1 for details).
These forecasts assume that the measurements are limited by statistical
rather than systematic errors, and much of the theoretical work
described above is aimed at understanding and correcting the potential
sources of systematic bias in each of the three structure probes.
Consistency among the different estimates of $w$ provides an external
check for systematics, but the real goal is to test for systematics
internally for each experiment and combine the results to obtain
higher precision and to search for departures from the simplest models.
The methods for combining results from different experiments are 
well understood in principle, but different issues arise depending
on what parameters one is trying to constrain.

One would ideally like to test for the presence of spatial curvature
rather than assume it to be zero, and in this case a
multi-probe approach like DES becomes much more powerful, because
curvature affects the distance-redshift relation but does not
affect the expansion history or the growth of fluctuations.
Another natural extension of the simplest models is a dependence
of the equation of state parameter on redshift, which can be
parameterized in various ways.  The most general approach is to
estimate $w(z)$ in bins and study the principal components of
the constraints for the various experiments, individually or in 
combination.$^{12}$ 
Probing to redshifts $z\sim 1$ with
high but not extraordinary (sub-percent) measurement precision, one can expect
to get a good constraint on an effective value of $w$
over the redshift range probed by the data,
and a loose constraint on redshift evolution of $w$.
One can take a similar approach using the Hubble parameter $H(z)$ or
the energy density $\rho_{\rm DE}(z)$ in place of $w(z)$.

The most exciting prospect of combining multiple probes is the 
possibility of distinguishing dark energy from alternative explanations
for cosmic acceleration, such modifications of General Relativity
or local inhomogeneities on horizon scales that cause apparent
accelerated expansion.
Regardless of the specific model, the hypothesis that some form
of dark energy drives cosmic acceleration entails particular 
connections between the expansion history, spatial geometry, and
the growth of fluctuations, and hence between the observables measured
by DES, by Planck, and their cross-correlation (the Integrated 
Sachs-Wolfe effect).  These connections can be different
in modified gravity models,$^{13}$ 
and perhaps in inhomogeneous universe
models as well.  Furthermore, even in some dark energy models the
clustering of dark energy can have an impact on CMB anisotropies
and matter clustering, and detecting these effects would give much
greater insight into the dark energy physics.  For each of these classes
of models, more theoretical work is needed to identify quantitative predictions
that distinguish them from simpler, uniform dark energy models and
to develop the best ways of testing these predictions.
The hope is that a quest for precise constraints on the 
dark energy equation of state will ultimately take us beyond $w$
to a deeper understanding of the origin of cosmic acceleration.

\medskip
\noindent
{\bf References}

\def\ref{\par\noindent\hangindent 15pt}
\parskip=1pt

\ref
1. The Dark Energy Survey, White Paper submitted to the
Dark Energy Task Force.  See also
{\tt http://www.darkenergysurvey.org/}.

\ref
2. Seo, H.-J., \& Eisenstein, D.~J.\ 2003, ApJ, 598, 720; 
Hu, W., \& Haiman, Z.\ 2003, Phys.Rev D, 68(6), 063004;
Glazebrook, K., \& Blake, C.\ 2005, ApJ, 631, 1 

\ref
3. Springel, V. et al.\ 2005, Nature, 435, 629;
White, M.\ 2005, {\tt astro-ph/0507307}

\ref
4. Ma, Zh., Hu, W., Huterer, D.\ 2005, ApJ in press ({\tt astro-ph/0506614})

\ref
5. Sheldon, E. S. et al.\ 2001, ApJ, 554, 881

\ref
6. Levine, E. S., Schulz, A. E., \& White, M.\ 2002, ApJ, 577, 569;
Majumdar, S., \& Mohr, J.\ 2003, ApJ 585, 603;
Hu, W.\ 2003, Phys. Rev. D,  67,  081304

\ref
7. Lima, M., \& Hu, W.\ 2005, Phys. Rev. D, 72, 043006;
Battye, R.A. \& Weller, J.\ 2003, Phys. Rev. D, 68, 083506

\ref
8. Cooray, A. \& Hu, W. 2001, ApJ, 554, 56; 
Zhan, H. \& Knox, L. 2004, ApJ, 616, L75;
White, M. 2004, Astroparticle Physics 22, 211

\ref
9. Jain, B. \& Taylor, A. 2003, PRL, 91, 141302;
Bernstein, G. \& Jain, B. 2004, ApJ, 600, 17;
Takada, M. \& Jain, B. 2004 MNRAS 348 897;
Hu, W. \& Jain, B. 2004, Phys.Rev. D70;
Zhang, J., Hui, L., \& Stebbins, A. {\tt  astro-ph/0312348};
Song PRD 71, 024026, 2005;
Knox, Song, Tyson {\tt astro-ph/0503644};
Bernstein, G. {\tt astro-ph/0503276};
Knox, L. {\tt astro-ph/0503405 }

\ref
10. Mandelbaum et al. {\tt astro-ph/0509026}

\ref
11. Huterer, D., Takada, M., Bernstein, B., and Jain,
B. {\tt astro-ph/0506030 };
Heymans, C. et. al. {\tt astro-ph/0506112};
Mandelbaum, R. et al. 2005, MNRAS, 361, 1287;
Jarvis, M. \& Jain, B. {\tt astro-ph/0412234}

\ref
12. Huterer, D. \& Starkman, G. 2003, PRL, 90, 031301

\ref
13. Lue, A. and Scoccimarro, R. and Starkman, G., PRD, 2004, 69, 044005;
Knox, L. and Song, Y. and Tyson, J.A., {\tt astro-ph/0503644}

\bye